\documentclass[times]{elsarticle}
\usepackage{graphicx}
\usepackage{subfig}
\usepackage{times}
\advance\voffset by -1.1in
\advance\hoffset by -0.9in
\begin{document}
\begin{frontmatter}
\title{Perception without Self-Matching in Conditional Tag Based Cooperation}
\author{David M. McAvity, Tristen Bristow, Eric Bunker, and Alex Dreyer \\
The Evergreen State College \\
         Evergreen Pkwy NW, Olympia, WA, 98505\\
           email: \texttt{mcavityd@evergreen.edu}}
\begin{abstract}
We consider a model for the evolution of cooperation in a population where individuals may have one of a number of different heritable and distinguishable markers or tags. Individuals interact with each of their neighbours on a square lattice by either cooperating by donating some benefit at a cost to themselves or defecting by doing nothing. The decision to cooperate or defect is contingent on each individual's perception of its interacting partner's tag. Unlike in other tag-based models individuals do not compare their own tag to that of their interaction partner. That is, there is no {\em self-matching}. When perception is perfect the cooperation rate is substantially higher than in the usual spatial prisoner's dilemma game when the cost of cooperation is high. The enhancement in cooperation is positively correlated with the number of different tags. The more diverse a population is the more cooperative it becomes. When individuals start with an inability to perceive tags the population evolves to a state where individuals gain at least partial perception. With some reproduction mechanisms perfect perception evolves, but with others the ability to perceive tags is imperfect.  We find that perception of tags evolves to lower levels when the cost of cooperation is higher.
\end{abstract}
\end{frontmatter}
\section{\bf Introduction}
\label{intro}
\vskip 5pt

One of the fascinating questions in evolutionary biology, which dates back to Darwin~\citep{darwin1859origin}, is how cooperation can emerge in a competitive environment where the struggle for survival and natural selection would seem to favour selfish behaviour. Since cooperation between individuals is widespread in nature~\citep{smith1997major}, from micro-organisms which exhibit complex social behavior~\citep{west2006social,crespi2001evolution} to social animals which form cooperative groups~\citep{clutton2000individual,sharp2005learned}, an explanation for its emergences is needed.  Among many theoretical approaches to this problem, evolutionary game theory~\citep{maynard-smith1974evolutionarygames},  in particular the prisoner's dilemma game, has proved to be fertile ground for research. One version of this game involves interacting individuals choosing one of two strategies: cooperate by donating a benefit to the other player at some cost to itself, or defect by offering nothing. An individual who defects will gain a fitness advantage compared to one who cooperates, but a group of cooperators who benefit from mutual cooperation is better off than a group of defectors who gain nothing. 

This observation points to a mechanism by which cooperative behaviour can evolve. If cooperative individuals preferentially interact with others who also cooperate with them they may gain a fitness benefit from the interaction. One way that this can happen is if the population  has some spatial structure or viscosity. The usual approach is to restrict individuals to particular locations on a spatial grid~\citep{nowak1992evolutionary,
nowak1994more,szabo1998evolutionary,schweitzer2002evolution,
langer2008spatial}. Alternatively individuals  may preferentially interact in certain isolated groups or demes, with limited migration between them, as in Wright's island model~\citep{wright1943isolation,taylor1992altruism,lehmann2006population,
rousset2000theoretical}. In a population with spatial structure individuals reproduce in their local neighbourhood, and therefore are more likely on average to have genes in common with their interacting partners. An individual who carries a gene for cooperation will preferentially donate benefits to other individuals who share that gene, and in turn receive benefits from those individuals. In this way the gene for cooperation can be sustained provided neighbours are sufficiently related.  The exact condition for the maintenance of cooperation depends on a number of factors, including the nature of the population structure, how costs and benefits relate to fitness, and the reproduction mechanism. In general, cooperative behaviour can survive and even grow provided initial clusters of cooperating individuals can form and the cost of cooperating is sufficiently low compared with the benefit given~\citep{langer2008spatial}. However, spatial structure alone may not be sufficient to sustain cooperation when the cost to benefit ratio is high. Indeed, for some update schemes, such as asynchronous birth-death schemes~\citep{huberman1993evolutionary,ohtsuki2006replicator,ohtsuki2008evolutionary}, spatial structure does not lead to the evolution of cooperation.

A more direct mechanism by which cooperating individuals may preferentially direct donations of a benefit to related individuals who cooperate with them is through kin recognition. Kin can be recognized through familiarity based on environmental or learned cues, or through pattern matching based on some inherited trait. If individuals display a heritable marker or  tag, such as a scent, color or some other phenotype, which interacting partners can recognize, individuals who preferentially cooperate with partners who share their own marker can gain an indirect benefit from that interaction, since those individuals are likely to be kin who also share the gene for cooperation.

A thought experiment by Hamilton~\citep{hamilton1964genetical} that illustrates how this mechanism can work even when individuals are not genealogical kin has come to be called the {\em green beard} effect~\citep{dawkins1976selfish}. If a gene for some distinguishable trait or tag (such as a green beard) also codes for a strategy of cooperating with others who share that tag then cooperation can be sustained.  Indeed, in the green beard effect cooperation will occur between any individuals who share the green beard gene, even if they are not genealogical kin. While a single gene or closely linked genes coding for tag, tag recognition and a cooperative response might seem too happenstance to be a likely biological model, there is some  evidence in support of this idea. For example, the social amoeba, {\em Dictyostelium discoideum}, which forms cooperative fruiting bodies, has been shown to preferentially direct benefits to carriers of the {\em csA} gene. Both the recognition and altruism behaviours derive directly from cell adhesion protein encoded by {em csA}~\citep{queller2003single}.

More generally genetic kin recognition refers a situation where individuals recognize and preferentially cooperate with other individuals based on some phenotype similarity, where the gene for recognizing and cooperating with related individuals and the gene for displaying the tag are different. Dawkins dubbed this the {\em armpit effect}~\citep{dawkins1982extended}. There is some question about whether genetic kin recognition can be sustained in nature due to what is sometimes referred to as Crozier's paradox~\citep{crozier}. If a successful genetic cue for cooperation leads to fixation and loss of diversity then this would open the way to exploitation by individuals who share the tag, but do not cooperate. Thus the success of genetic kin recognition as a route to cooperation is sensitive to the maintenance of of diversity in tags through mutation or some other mechanism~\citep{rousset2007constraints,jansen2006altruism,gardner2007social}.

Several theoretical models for the evolution of cooperation involving heritable markers or tags have been proposed to explore this question~\citep{riolo2001evolution,sigmund2001tides,traulsen2003minimal,axelrod2004altruism,jansen2006altruism,hammond2006evolution,crozier,rousset2007constraints}. The mechanism by which tags are introduced varies, but the common thread is that individuals adopt strategies that are contingent on the tag of their opponent. Both the tag and the strategy are inherited and are subject to mutation, although they need not evolve together.  Riolo \textsl{et al.}~\citep{riolo2001evolution} and Sigmund \textsl{et al.}~\citep{sigmund2001tides} discuss a model in which well-mixed individuals cooperate with those who have a tag that is within a particular tolerance level of their own, and defect otherwise. The tags in this instance are continuous. The system can evolve to a state with a relatively high level of cooperation, were most individuals share similar tags, but have a relatively low tolerance level.  However the dynamics are quite unstable, leading to "tides of
tolerance". Highly intolerant mutant strategies typically invade more tolerant and cooperative ones, which results in a drop in the cooperation rate. Eventually mutations give rise to more tolerant individuals with a different tag  -- resulting in a return to cooperative behaviour, and the cycle continues. Traulsen and Schuster~\citep{traulsen2003minimal}, proposed a discrete version of this model, in which there are two tags and two levels of tolerance. This model is amenable to analysis using replicator dynamics and leads to a similar dynamical situation. Roberts and Sheratt~\citep{roberts2002Similarity} pointed out that in these models  interacting individuals who share identical tags unconditionally cooperate, as in the green beard effect, so it is not surprising that cooperation evolves.  When they made a modification that allows individuals the option of not cooperating with those who have identical tags the results was a loss of cooperative behaviour. However, it has since been shown that if the mutation rate for tags is greater than the mutation rate for strategies, cooperative behaviour can predominate, even if like individuals are not assumed to cooperate with each other~\citep{traulsen2007chromodynamics}. 

Another approach has been to study these and more general tag models for the case where individuals are not uniformly mixed, but are constrained by some viscosity or spatial structure~\citep{jansen2006altruism,hammond2006evolution,crozier,
rousset2007constraints}. While cooperation can evolve in the spatial prisoner's dilemma without tags provided the cost of cooperation is low compared to the benefit and the update scheme is favourable, the inclusion of conditional cooperation based on heritable tags can enhance the rate of cooperation and may allow it to evolve where it would not otherwise do so. In the model introduced by Jansen and Van Baalen~\citep{jansen2006altruism}, individuals can adopt a strategy of either cooperating or defecting against those who share their tag and defecting against those with a different tag. The dynamics is relatively stable and cooperative provided the tag and strategy are not always inherited together. One interesting feature is the positive correlation between number of tags and the level of cooperation. Hammond and Axelrod~\citep{hammond2006evolution} allowed for the additional conditional strategy that individuals could optionally cooperate with those who had a dissimilar tag. Although such behaviour rarely evolved, they showed that the cooperation rate is sustained at levels above what is normally expected in the spatial prisoner's dilemma even when the cost of cooperation is high.

Thus a common outcome in tag models with spatial structure or viscosity is that cooperation is enhanced by the presence of heritable tags in a population -- with the nice result that tag diversity yields a higher cooperation rate. However, while individuals cooperate within groups of the same tag, they almost invariably defect against those with different tags. We refer to cooperation with liked-tagged individuals as {\em loyalty} and cooperation with dissimilar others as {\em  hospitality}. Individuals who are loyal but in-hospitable are said to exhibit {\em ethnocentrism}~\citep{hammond2006ethnocentrism} or {\em nepotism}, and this is the typical way that cooperation manifests itself in these models. Thus defection between individuals is now replaced by defection between groups of otherwise cooperative individuals. While this may seem unfortunate from a social perspective, the overall rate of cooperation is higher as a result. From a biological perspective preferentially donating to kin seems an eminently reasonable strategy for propagating ones genes and exemplifies Hamilton's idea of inclusive fitness~\citep{hamilton1964genetical}.

The models mentioned above assume a form of kin discrimination based on {\em self-matching}, i.e. kin recognition by comparing tags of others with the tag of oneself. While this is plausible biological mechanism with some evidence to support it~\cite{mateo2000kin}, kin recognition based on learned or environmental cues due to prior association such as during rearing seems more plausible. Individuals with limited cognitive ability may not have the ability to make a comparison between themselves and their interacting partner; they may not even have the self-awareness to recognize their own tag. Individuals might simply use the observed tag of a partner as a cue or trigger for a cooperative or defecting response. With this in mind, we introduce a class of tag based contingent strategies that do not involve an individual comparing its own tag to that of the individual it is interacting with. Instead individuals opt to cooperate or not based solely on their perception of their partner's tag. 

We also wish to explore the effect of imperfect perception of tags on the evolution of cooperation and the extent to which the perception of tags can evolve. To this end we introduce the possibility that individuals are occasionally unable to recognize the tag of an opponent due to some limitations in their perception. Limitations in perception may be due to some exogenous environmental condition, such as darkness or fog, or it could be endogenous, with some individuals having a better ability to perceive tags than others. We analyse both situations.

\section{\bf The Model}
\label{model}
We consider spatially separated individuals, each inhabiting a single cell on a square lattice, playing the one shot prisoner's dilemma game with the eight adjacent individuals in their Moore neighbourhood. The environment is saturated, with no empty cells. We use a common implementation of the prisoner's dilemma game, which involves interactions between two individuals, who can optionally offer some benefit $b$ to their opponent at some cost $c$ to themselves. Individuals who offer a benefit are said to cooperate, and those who do not defect. In any single interaction individuals always stand to gain more if they defect then if they cooperate. However, a group of mutual cooperators, who gain an average payoff of $b-c$ ( $c<b$) are better off than a group of mutual defectors, who gain nothing.

Each individual has one of $m$ possible heritable tags. The tag is represented by an integer, $\ell\in [m]=\{1,2,\cdots,m\}$ and it forms part of a genome which also includes an array which codes for a conditional strategy. Each entry in the array is the probability of cooperating with an individual with a tag number corresponding to the array index. For example, if there are four tags, then the genome will be a tag number between 1 and 4 and a strategy array containing four probabilities. Two different types of models are possible. In the first, with pure strategies, probabilities are either 0 or 1. This means that an individual will either always cooperate (1) or always defect (0) when encountering another individual with a particular tag.  An individual with strategy array [1,1,0,1], for example, cooperates with all individuals except those with tag 3. In the second case, with mixed strategies, the probability of cooperating with an individual with a particular tag can be any real number on the interval $[0,1]$. In the pure strategy case the genome is a member of the strategy set $[m]\times\{0,1\}^m$. The mixed strategy case the genome is a member of the strategy set $[m]\times [0,1]^m$. In this paper we deal with the pure strategy case and will consider the mixed strategy case in a future paper. 

Individuals start with randomly assigned strategies and tags. Each individual, $i$, plays the prisoner's dilemma game once with each of its eight neighbours, accumulating a total pay-off $\pi_i$, based on any costs it incurs or benefits it gains. Individuals reproduce asexually via the Moran process, whereby an individual is targeted for death and then it and all its neighbours compete to replicate in the cell that it will vacate. In synchronous updating all agents are are targeted for death and then are replaced by replication at each time step. In asynchronous updating only a single agent is replaced each time step. We consider both approaches.

The probability that an individual is successful in replicating in a vacated cell is proportional to its fecundity relative to the fecundity of all cells neighbouring that vacant cell. Fecundity, $f_i$ is assumed to be an increasing function of the pay-off $\pi_i$. For an individual $i$ with fecundity $f_i$  competing to reproduce in a cell, the probability of success, $p_i$ is 
$$p_i={f_i\over \sum_{j\in\cal N} f_j}\; ,$$
where the sum in the denominator is over the set, ${\cal N}$, of all individuals neighbouring a vacated cell including the occupant of the cell that will be replaced. If $\sum_{j\in\cal N} f_j=0 $, then the probability for replication is equal for all individuals. Two classes of function for the fecundity are common in the literature. In the exponential or Boltzmann model~\citep{szabo1998evolutionary,szabo2002phase,traulsen2006stochastic}
$$f_i=e^{\beta \pi_i}\, ,$$
where $\beta$ is the intensity of selection. 
The probability of replication for individual $i$ becomes
$$p_i={1\over 1 + \sum_{j\in {\cal N}|_{ j\ne i}} e^{\beta (\pi_j-\pi_i)}}\, .$$
In this model only the relative payoffs are important in determining the probability of replication. For $\beta \to 0$ all individuals are equally likely to replicated, which corresponds to neutral drift. The limit $\beta \to \infty$ corresponds to a type of strong selection where the individual with the highest payoff competing for a vacated cell is always chosen to replicate. This method of updating is not uncommon~\citep{nowak1992evolutionary,ohtsuki2008evolutionary,hauert2001fundamental,killingback1996spatial} and is sometimes called {\em imitation} because it corresponds to an individual imitating the behaviour of the neighbour with the highest pay-off. 

Another common model is to assume that fecundity is a linear function of the payoff~\citep{ohtsuki2006replicator,axelrod2004altruism,riolo2001evolution}.
$$f_i=1+\beta \pi_i$$
In this model the $\beta\to 0$ limit corresponds to neutral drift and the $\beta \to \infty$ limit gives a different, more  common, definition of strong selection, where the probability of replication is directly proportional to an individual's payoff. We call this proportional updating, for which the probability of replication for individual $i$ becomes
$$p_i={\pi_i\over \sum_{j\in \cal N} \pi_j}\; .$$ 
While these models differ in their interpretation of strong selection they correspond in the small $\beta$ or weak selection limit. Because prior results in the spatial prisoners dilemma games show some sensitivity to the nature of the update scheme, we consider synchronous and asynchronous updating and both the Boltzmann and linear fecundity functions. In most cases the results are qualitatively similar, but there are differences which we will discuss when they arise. Unless otherwise stated the results quoted correspond to the synchronous updating using imitation, i.e. the strong selection version of the Boltzmann  model. We choose this as our baseline for comparison, since this was the model for early work in the spatial prisoners dilemma, and because as a deterministic model it lends itself to some straightforward quantitative analysis that helps explain some of the emergent behaviour.

In addition to the replication mechanism above, we also take into account mutation. When an individual is selected to replicate in a vacated cell there is a  probability $\mu$ of mutation. Each part of the genetic code mutates independently, including the individual bits of the strategy array. When the tag gene mutates a random tag is assigned. Each item in the strategy array mutates independently with $0 \to 1$ or $1 \to  0$ in the event that a mutation occurs.

Our objective is to determine what types of strategies evolve in a simulation of this game and, in particular, to determine the proportion of individuals who engage in cooperative behaviour. We call individuals who cooperate with any individual, regardless of tag, All-C. Individuals who defect against all other are called All-D. Individuals who only cooperate with other individuals who share their tag and defect against all other types of individuals are called {\em ethnocentric} individuals, as indicated earlier. Many other strategies exist.

The simulations were run using NETLOGO 4.1 on a 200x200 toroidal grid for a total of 40,000 individuals. Simulations were run long enough for the system to reach a stable cooperation rate, if one existed. This was typically achieved within 1000 generations in the case of no mutations, and as long as 10000 generations when considering mutations. Longer runs were necessary when the cost to benefit ration $c/b$ was at a transition value between highly cooperative behaviour and highly defective behaviour. In these cases there was not always a clear or unique stable cooperation rate. Four quantities were recorded in each run: The mean cooperation rate, the diversity of individuals, the mean probability of cooperating with like-tagged individuals (loyalty), and the mean probability of cooperating with different tagged individuals (hospitality). The cooperation rate is the proportion of individuals who cooperate each generation and is equal to the mean fitness of the individuals as a proportion of the fitness that would result if all individuals cooperated. For a measure of diversity  Simpson's diversity index was used
$$ D={N(N-1)\over {\displaystyle \sum_{\ell=1}^m }n_\ell (n_\ell-1)}$$
where $N$ is the total number of agents, $m$ is the number of tags and $n_\ell$ is the number of individuals with tag $\ell$. For large $N$ the maximum value of Simpson's diversity index $D$ approaches the number of tags $m$. 

We define perception $\rho \in [0,1]$ as the probability that an individual is able to identify the tag of its opponent. If an individual is unable to identify the tag of its opponent it randomly assigns a tag and then follows the appropriate strategy for that tag. An alternative approach would be to have individuals employ a separate strategy for the case when they do not recognize a tag. We do not pursue this alternative here.   We first consider the case where perception is exogenous. All individuals have the same perception that does not change. Then we consider the case of endogenous perception, whereby each individual has its own heritable level of perception which is also subject to mutation.

\section{Results and Discussion}
\label{results}
\subsection{Perfect Perception}
\label{perfect}
We first establish how the cooperation rate depends on (a) the cost to benefit ratio and (b) the number of tags, when there is no mutation and perfect perception. With synchronous updating using the strong selection limit of the Boltzmann fecundity function this model is deterministic.

 The cost to benefit ratio was varied between 0.1 and 0.9. This procedure was repeated for each of the tag numbers 1, 2, 4 and 8.  With tag number set to 1, so there are no distinguishable tags,   the simulation should be consistent with the results of the usual spatial prisoner's game~\citep{nowak1992evolutionary,nowak1994more,szabo1998evolutionary,schweitzer2002evolution,langer2008spatial}.

\begin{figure}[!h]%
\centering
\includegraphics[width=3.5in]{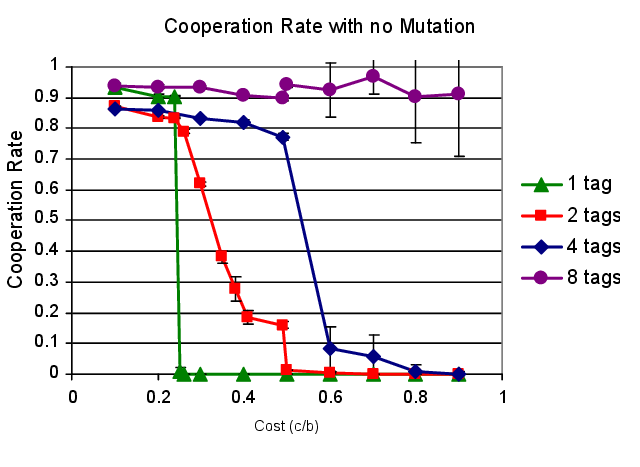}%
\caption{Mean cooperation rate for 10 replicates is shown as a function of the cost to benefit ratio for different numbers of tags, with error bars showing the standard error. There is a high cooperation rate for $c/b< 0.25$ regardless of the number of tags. The cooperation rate decreases with increasing cost for $0.25<c/b<0.5$, with cooperation sustained at higher rates at higher costs when there are more tags. For $c/b>0.5$ the cooperation rate drops to zero, except for the case of 8 tags when the cooperation rate is high on average, but highly variable. }%
\label{fig:fitness1}%
\end{figure}

Cooperation rate as a function of cost is shown in figure~\ref{fig:fitness1}, with the mean of ten replicates plotted and error bars indicating the standard error.

In all cases the cooperation rate is high (approximately 0.9) for $c/b < 0.25$, with the cooperation rate for 2 and 4 tags being somewhat below the case for 1 tag and 8 tags. Thus, the presence of distinguishable tags does not significantly increase cooperation for $c/b<0.25$. In this region of parameter space individuals form like-tagged clusters employing an ethnocentric strategy. Individuals that survive cooperate only with other individuals who share their own tag, even though they do not self-match. Defection occurs on the boundary between two ethnocentric clusters with different tags, or due to the presence of small numbers of All-D individuals on the boundary of a like tagged cluster. This result is not unexpected, since in the model with 1 tag, were there are no conditional strategies, cooperators dominate for these costs.  When conditional strategies based on distinguishable tags are included ethnocentric strategies will be fitter than All-C strategies, because they will be more effective at eliminating rare All-D strategies in their clusters. 

\begin{figure}[!h]%
\centering
\subfloat[]{\includegraphics[width=2 in]{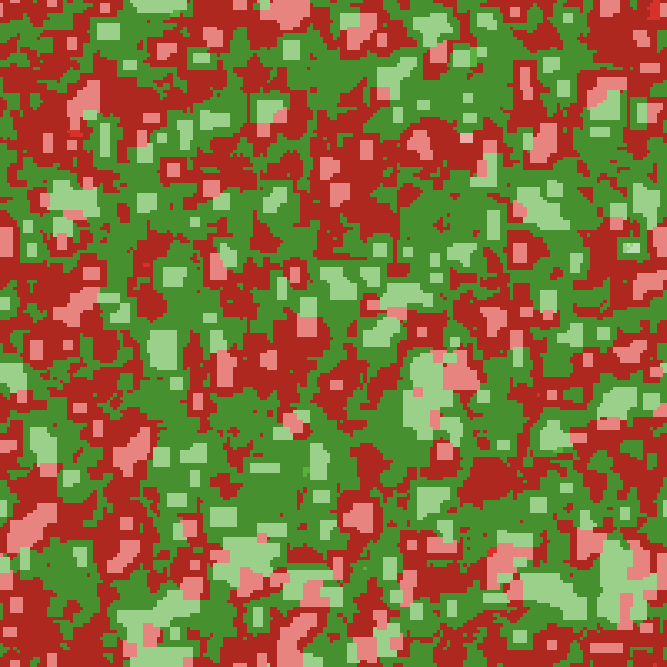}}\quad%
\subfloat[]{\includegraphics[width=2 in]{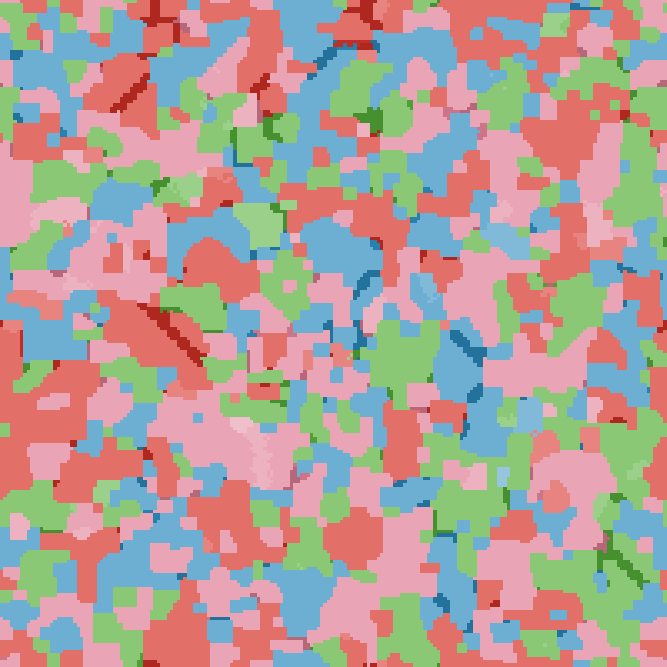}}\quad%
\subfloat[]{\includegraphics[width=2 in]{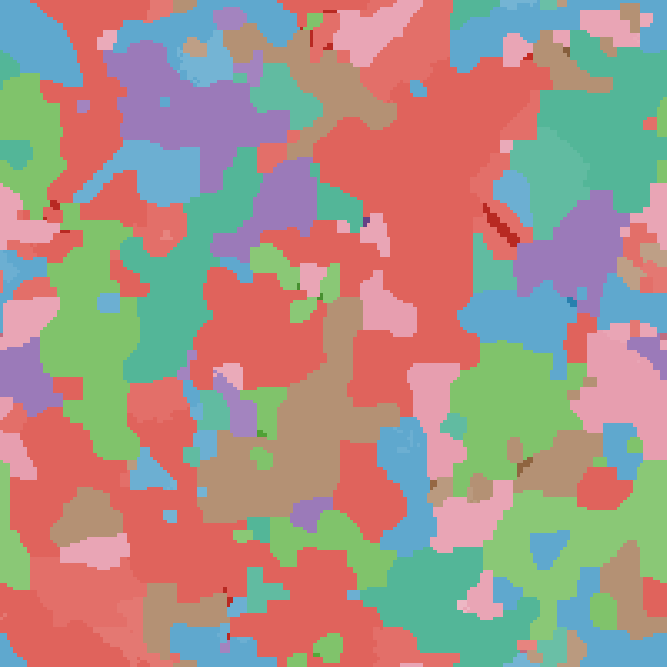}}%
\caption{World view for $c/b=0.4$ for two tags (a); four tags (b); and eight tags (c). Colours label the tags, with lighter shades indicating more cooperative behaviour. The cooperation rate increases with increasing tag number. The size of clusters of like-tagged individuals also increase with increasing number of tags.}%
\label{fig:worlds}%
\end{figure}

For $0.25<c/b<0.5$ there is a drop in the cooperation rate with increasing cost. The drop is most pronounced when the number of tags is small. As shown in the appendix, this transition is a result of the fact that All-D clusters are able to resist invasion by cooperative individuals for $c/b>1/4$. For one tag,  3 by 3 clusters of cooperators can only invade defectors for $c/b<1/4$, although they are stable against invasion by defectors for $c/b<5/8$. Thus for $1/4<c/b<5/8$ the world becomes mostly populated by defectors, with only small islands of stable cooperators. As a consequence, the cooperation rate is very low. For a more complete discussion of fundamental clusters for a variety of spatial games see Hauert~\citep{hauert2001fundamental} and earlier work by Nowak {\em et. al.}~\citep{nowak1992evolutionary,
nowak1994more}. With 2 or more heritable tags, All-D individuals dominate at the start of the simulation for $c/b>1/4$, causing the cooperation rate to fall. However, as these individuals encounter ethnocentric individuals with different tags, their invasion is halted and reversed. A detailed analysis of different configurations of clusters with two tag types is given in the appendix and shows that we should expect a transition from cooperative behaviour to non-cooperative behaviour to occur somewhere in the interval $2/5<c/b<1/2$. This is consistent with what is observed. The more types of tags there are the more likely All-D individuals are to find themselves interacting with ethnocentric individuals with different tags. Thus the cooperation rate is higher the more tags there are.  We see a positive correlation between cooperation rate and number of tags, at least for these levels of cost. Jansen {\em et al.}~\citep{jansen2006altruism} observed a similar result in their model. Figure~\ref{fig:worlds} shows the results of three simulations for the case where the cost to benefit ratio is 0.4. Figure~\ref{fig:worlds}(a) shows the case of two tags (coloured red and green). The darker shades are defecting strategies, and the lighter shades are ethnocentric strategies, cooperating with their own tag, but not the others. Figure~\ref{fig:worlds}(b) has four tags with colors red, green, purple and blue. The proportion of light shaded ethnocentric strategies is higher, indicating the higher levels of cooperation. Figure~\ref{fig:worlds}(c) shows the case with eight tags, with a similar highly cooperative state. Notice that the cluster sizes of each tag are larger in the eight tag case, and that almost all the defecting strategies have been eliminated. Because cluster size for the eight tag case grows so large, the diversity index tends to drop well below the maximum value of 8. When simulations are run with a larger world size the diversity does not drop as much, yet the cooperation rate remains in the same range.

There is a clear transition in behaviour of the models at $c/b=0.5$. For $c/b<0.5$ the cooperation rate reaches a steady state in less than 200 generations, and is consistent from run to run. For $c/b > 0.5$ the behaviour of the model is highly unpredictable for 4 and 8 tags. For these high costs, defecting clusters are able to invade cooperators at a rate that is faster than they are invaded by ethnocentric strategies with unlike tags. However, if All-D individuals eliminate a local population of like tagged cooperating individuals, they will eventually succumb to the ethnocentric individuals that are invading them. The dynamics can be quite involved. Occasionally ethnocentric individuals fixate, at other times All-D individuals do. Often, it takes over 5000 generations before fixation occurs, with the ultimate fate frequently being decided by some chance encounter near the end of the simulation. For this reason we ran 30 replicates of these simulations. Cluster size grows and diversity drops as the simulation runs. For 8 tags fixation of ethnocentric strategies is the most likely outcome, even for very high cost. For 4 tags fixation of All-D strategies becomes increasingly likely as cost increases.

There is another outcome that can arise for $c/b > 0.5$. Mutual coexistence of individuals with two different tags can evolve, with the following tag-strategy combination: individuals defect against other individuals who share their own tag and cooperate with individuals of the other tag type. Such individuals are disloyal, but hospitable. They could be called traitors~\citep{shultz2008stages}, although this term does not adequately describe the new form of cooperative behaviour that results. Rather than forming well defined clusters, these individuals form an intricate web of the two different tag types with mutual support that helps them resist invasion by most other strategies. The equilibrium is dynamic, with individuals alternating strategies and tags every other generation. While this situation exhibits hospitality it comes with a lack of loyalty. The overall rate of cooperation of such configurations is consistently less than what would result with ethnocentric individuals, but is higher than would occur if all individuals were All-D. This type of behaviour seems to be an artefact of the deterministic updating process. With asynchronous updating such behaviour is only transitory.

\subsection{Mutation}
\label{mutation}
\begin{figure}[!h]
\centering
\includegraphics[width=3.5in]{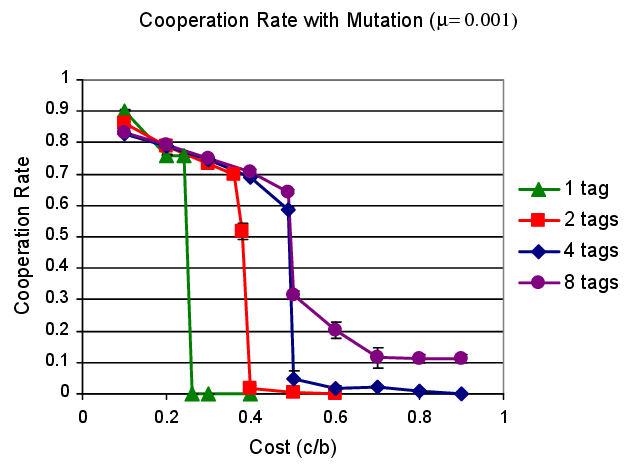}%
\caption{Mean cooperation for 10 replicates is plotted as a function of cost to benefit ratio with the mutation rate $\mu=0.001$. Error bars, when visible, indicate standard error. As $c/b$ increases from 0.1 there is a gradual decrease in the cooperation rate until a sharp transition to zero or low cooperation. This transition occurs at $c/b=0.25$ for 1 tag, at $c/b=0.4$ for 2 tags and $c/b=0.5$ for 4 and 8 tags.  For 8 tags the cooperation rate does not fall to zero for high cost, but approaches a cooperation rate of approximately 0.1. The high rate of diversity introduced by mutation allows a small fraction of ethnocentric strategies to be maintained in dynamic equilibrium with the more dominant defectors. }
\label{fig:fitness2}%
\end{figure}

In the above results there was no mutation.  With a mutation rate $\mu=0.001$ per generation, per gene, the results are qualitatively similar to the case with no mutation, although the overall level of cooperation is reduced for all costs. The complex dynamics for $c/b>0.5$ also disappears. The results are shown in figure~\ref{fig:fitness2}.

As $c/b$ increases from 0.1 there is a gradual decrease in the cooperation rate until a sharp transition to zero or low cooperation. This transition occurs at $c/\approx0.25$ for 1 tag, at $c/b\approx0.4$ for 2 tags and $c/b\approx0.5$ for 4 and 8 tags. These transitions are consistent with the cluster analysis in the appendix. For 8 tags the cooperation rate does not fall to zero for high cost, but approaches a cooperation rate of approximately 0.1. Thus the consequence of mutation when there are 8 tags is to undermine the highly cooperative behaviour for $c/b > 0.5$ that was sustained as a result of complex dynamics with no mutation. This low but non-zero cooperation rate for 8 tags is sustained by the balance between the weak selection pressure in favour of All-D individuals over ethnocentric ones at high costs and mutation, which introduces new tags and ethnocentric strategies which prevent the All-D individuals from fixating.

It is worth comparing the results above to those that come about from the other update  mechanisms discussed in the previous section. While there is little difference between the synchronous and asynchronous updating with mutation, there are some differences between the imitation updating function discussed above, and the proportional updating based on the strong selection limit of the linear fecundity function. The key difference is that the transition to no cooperation at low cost for 1-tag occurs earlier, at $c/b\approx 0.125$ in asynchronous proportional updating. This fits with previous analysis of spatial prisoner's dilemma which show that for this type of updating, cooperative behaviour can only be sustained if $c/b< 1/n$, where $n$ is the number of neighbours of an interacting cell~\citep{ohtsuki2006simple}. In our model $n=8$. When there are two tags, the transition to low cooperation again occurs at $c/b\approx 0.4$. For 4 and 8 tags there is not a sharp transition to low cooperation rates, and cooperation rates are maintained at higher levels at high costs than in the imitation updating.  An explanation for this difference is that at high costs All-D individuals are able to invade more successfully with imitation updating than with proportional updating, consequently mutations that introduce new tags with ethnocentric strategies are able to maintain the level of cooperation at higher levels.

\subsection{Exogenous Perception}
\label{globalperception}
\begin{figure}[!h]
\centering
\includegraphics[width=3.5in]{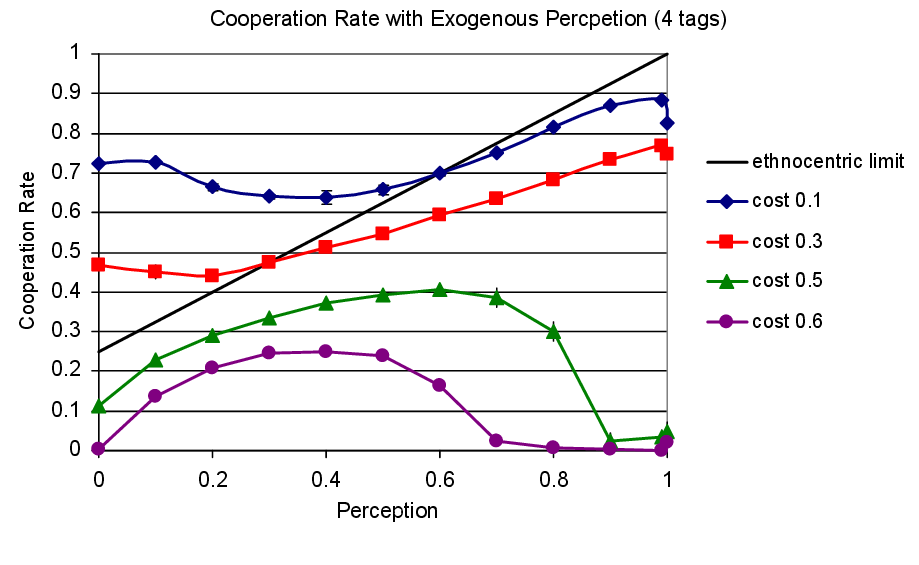}%
\caption{The mean cooperation rate for 5 replicates is plotted as a function of  perception for different cost to benefit ratios. Error bars, where visible, are for standard errors. Mutation rate is 0.001. Cooperation rate is higher for lower costs, as expected. As perception increases the cooperation rate curves approaches a line we call the ethnocentric limit. This line gives the expected cooperation rate if all individuals are ethnocentric with the same tag. Cooperation rates initially decrease with increasing perception when cost is low, but increase when cost is high. Cooperation rates are highest when perception is near perfect for $c/b=0.1$ and $c/b=0.3$, but achieves its maximum value for imperfect perception for $c/b=0.5$ and $c/b=0.6$.}%
\label{fig:perception}%
\end{figure}

We now introduce the possibility that individuals have a limited perception of their interacting partner's tag.  In this section all individuals have the same exogenous perception, which corresponds to a situation where individuals have their ability to perceive tags limited by an environmental constraint of some kind. We restrict our attention to the 4 tag model, although results for 2 and 8 tags are similar. The mutation rate is fixed at $\mu=0.001$. We vary perception between $\rho=0$ (no ability to perceive tags) and $\rho=1$ (perfect ability to perceive tags), for a variety of different costs. The results are shown in figure~\ref{fig:perception}, with cooperation rate plotted vs. perception for costs 0.1, 0.3, 0.5 and 0.6.
Consider the left axis of the figure, where perception is zero. For each cost the cooperation rate is lower than the case of perfect perception which we considered earlier and is shown on the right axis. When $\rho=0$ individuals associate a random tag to all opponents, completely independent of the actual tag. Hence the benefits of heritable tags associated with our earlier analysis are diminished. Individuals choose to cooperate or defect depending on a randomly assigned tag and the corresponding entry in their particular strategy array. For example, All-C individuals with a strategy array with four cooperating genes [1,1,1,1] would still cooperate all the time, but an individual with an array with three cooperating genes, such as [1,0,1,1],  would end up cooperating 75\% of the time and defecting 25\% of the time. For cost 0.1, the equilibrium cooperation rate is close to 0.75, indicating that generally cooperative individuals with three cooperating genes and one defecting gene fixate. Such individuals are loyal to their own tag and hospitable with at least some other tags. As cost increases, cooperating genes become less and less favourable until All-D fixates at cost 0.6.

As perception increases the cooperation rate approaches a common limiting line that we have called the ethnocentric limit. This line shows the cooperation rate of individuals in a world where all individuals have the same tag type and follow an ethnocentric strategy. In such a world, the cooperation rate $r$ is given by the expression:
$$r=\rho+(1-\rho){1\over m}={m-1\over m}\rho+{1\over m}$$
where $\rho$ is the level of perception and $m$ is the number of tags. Points above this line indicate the presence of strategies that are more cooperative than ethnocentric strategies. Points below this line indicate the presence of more All-D strategies, or smaller clusters of like-tagged individuals with ethnocentric strategies. As perception increases from $\rho=0$ it becomes increasingly favourable for individuals to adopt an ethnocentric strategy. For certain levels of perception the world becomes populated with large clusters of ethnocentric strategies and the cooperation rate approaches and then tracks the ethnocentric limit. However, as perception continues to increase, the cooperation rate falls below the ethnocentric limit. This is chiefly because ethnocentric clusters become smaller and thus the number of interactions with unlike tagged opponents on the boundary between clusters becomes more common. In addition, as perception improves, All-D individuals are increasingly able to dominate when the cost of cooperation is high.

For each cost there is a level of perception where the cooperation rate is maximum. For $c/b= 0.1$ and $c/b= 0.3$ the maximum is at a perception close to 1.0. For $c/b= 0.5$ the maximum cooperation rate occurs at perception 0.6, and for $c/b= 0.6$ the maximum is at perception of about 0.4. We conclude that imperfect perception of inherited tags can enhance cooperation significantly compared to the case with perfect perception when the cost to benefit ratio is high. The question that arises is, if each individual had its own inherited level of perception which was subject to mutation, would perception evolve to a level that corresponds to a maximum cooperation rate at each cost? We pursue this question next.

\subsection{Evolving Endogenous Perception }
\begin{figure}[!h]
\centering
\includegraphics[width=4.2in]{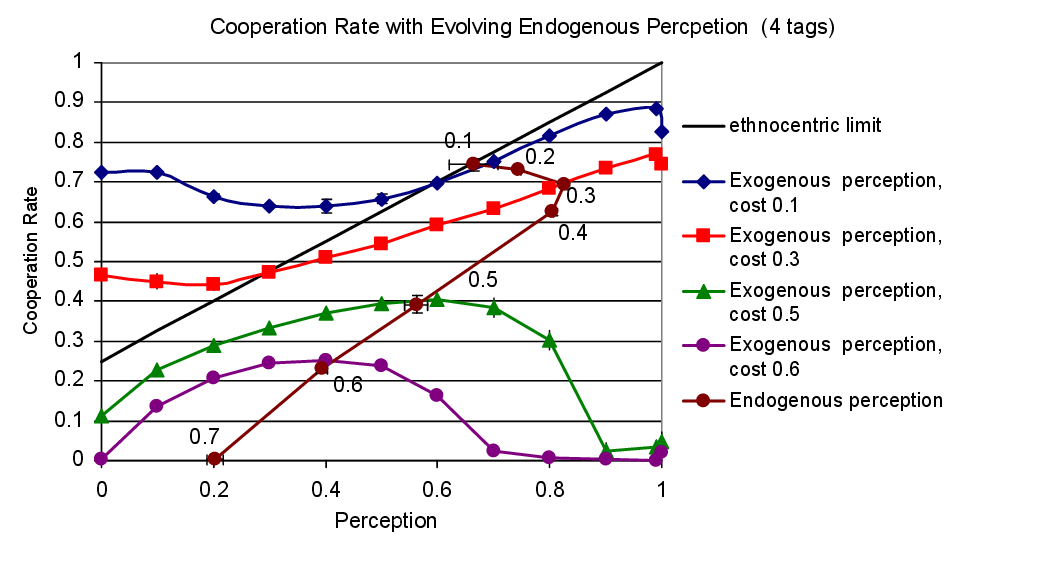}%
\caption{Mean cooperation rate for 5 replicates is plotted as a function of perception with mutation rate $\mu=0.001$. Error bars, where visible, show the standard error. The evolved perception line shows the mean cooperation rate and the mean perception levels that resulted when perception was allowed to evolve with different cost to benefit ratios. The corresponding cost to benefit ratios are labelled on the graph. This line is superimposed on the cooperation rate versus perception curves for fixed exogenous perception. Perfect perception does not evolve. For high cost to benefit ratio perception evolves to a level that maximizes the cooperation rate. }%
\label{fig:localperception}%
\end{figure}

We now consider the case where perception is an endogenous variable, so that each individual has its own heritable level of perception. Perception thus becomes another part of each individual's genetic code which is allowed to mutate independently. When the perception gene mutates the perception changes by an amount drawn from the normal distribution with mean 0 and standard deviation 0.1. Perception is prevented from going below 0 or above 1 by rescaling. We choose a mutation rate of 0.001, as before. In what follows all individuals initially have zero perception. The level of perception evolves to non-zero values through mutation and selection. Similar results apply for other initial conditions, such as when each individual is randomly assigned an initial perception between 0 and 1.  The simulation was run for costs varying from 0.1 to 0.9 for up to 10,000 generations until the cooperation rate and the level of perception reached a steady state. The results are shown in figure~\ref{fig:localperception}. The relationship  between the cooperation rate and the corresponding evolved perception levels are superimposed on the graph from the previous section figure~\ref{fig:perception}. 

The evolved perception curve shows that for high cost, $c/b\ge 0.5$,  perception evolves to a level that maximizes the rate of cooperation. For $c/b= 0.6$ the perception evolves to 0.4, with a resulting cooperation rate of 0.25. For $c/b= 0.5$ the perception evolves to 0.5 resulting in a cooperation rate of 0.4. In both cases the cooperation rate is much higher than it would be for perfect perception, when it is approximately zero. For lower costs, the optimal level of cooperation would occur for perception near 1. However, for low costs perception evolves to between 0.7 and 0.8, resulting in suboptimal rates of cooperation. This can be understood as follows. With perfect perception at low cost, ethnocentric strategies predominate, even with mutation. However, if an ethnocentric individual in a cluster of like tagged ethnocentric individuals has slightly lower perception than the average individual in its neighbourhood it will benefit from the occasional defection that results from misperceiving a neighbour's tag. Hence, lower perception is selected for. However, if the perception is too low, such individuals will be at a disadvantage when they encounter unlike-tagged ethnocentric individuals at the boundary of a cluster, who will not cooperate with them, but with whom they will cooperate occasionally. Thus a relatively high perception that is less than perfect represents the equilibrium that balances these competing selection pressures. This is sensitive to the  number of cells in the middle of a cluster relative to the number that are on the boundary.

With asynchronous imitation updating the results are similar to the above, however, with asynchronous proportional updating we see an interesting difference. In this case the highest cooperation rates for exogenous perception occur when perception is perfect, regardless of the cost. In addition, when perception is endogenous, populations that start with zero perception evolve to a state of near perfect perception when the cost is less than 0.7. For costs higher than this there is no cooperation and perception is not selected for. We can account for this difference between the two update schemes as follows. At low cost, clusters of like-tagged ethnocentric strategies form as with imitation updating. However, proportional updating leads to more interactions between individuals of different tags at the border between clusters because the fittest individuals are not always successful at reproducing in vacated cells as they are in imitation updating. Consequently the advantages of cheating by misperceiving within a cluster of like tagged individuals is outweighed by the disadvantage of misperceiving unlike tagged individuals on the border of the cluster.

\section{Conclusion}
The results demonstrate that the presence of heritable tags with strategies for cooperation that are contingent on the tag of an opponent leads to a significant increase in overall cooperation in the spatial prisoner's dilemma game, compared to the case when there are no tags, so long as $c/b>0.25$. The most common individuals are those with an ethnocentric strategy, whereby individuals cooperate with like-tagged individuals and defect against individuals with different tags. This happens despite the fact that no explicit self-matching occurs.  High levels of cooperation can be maintained for increasingly large costs as the number of tags increases, particularly for asynchronous proportional updating.

Interestingly, when individuals occasionally misperceive the tag of other individuals, cooperation can be further enhanced when the cost is high ( $c/b\ge0.5$). With imitation updating, when the ability to perceive is allowed to evolve, individuals evolve partial but less than perfect perception. For high cost of cooperation this results in individuals achieving an even higher rate of cooperation than they would for perfect perception of tags.  Another interesting consequence of partial perception is that such individuals will occasionally cooperate with individuals with unlike tags.

In this paper we have proposed one possible model for perception, for which the results prove interesting. However, it is certainly not a universal model. For example, in this implementation when an individual misperceives the tag of its interacting partner  it assigns a random tag and then adopts an existing strategy for that tag. An alternative mechanism would be to have a separate strategy that is adopted any time a tag is not recognized. No doubt other models are possible. More generally, while we see an increase in the cooperation rate due to conditional tag-based strategies, these strategies are contingent on the ability to perceive tags, and this ability might reasonably have a fitness cost. Our model as implemented does not account for costly perception. Preliminary work suggests that if the cost of perception is sufficiently small, perception still evolves, but that perception fails to evolve for sufficiently high cost. Finally, we use the term "perception" for the probability that an individual 
identifies the tag of its interacting partner and responds with the correct strategy for that tag. However, since the simulation does not include a physical perception mechanism, other interpretations of this probability are possible. For example, imperfect endogenous perception could be interpreted as an occasional disinclination to pay attention to tags when making a choice about whether or not to cooperate, rather than simply an inability to identify a tag correctly. Perception would thus be a measure of sensitivity to tags.  Alternatively, the perception probability may be interpreted simply as a mechanism for individuals to partially mix the pure strategies in their genome in a constrained way. As mentioned earlier, imperfect perception allows ethnocentric strategies to "cheat" by occasionally defecting when interacting with similar others. A natural question to ask is what would happen to  the cooperation rate if individuals could mix strategies in an unconstrained way and how would perception evolve in such a model. This is a topic that will be explored more fully on both analytical and agent-based levels in a future paper that explores other natural mechanisms for mixing strategies.

The model, as given, demonstrates results that should be testable. When the cost of cooperation is low the perception of distinguishable tags will be high and the level of ethnocentric behaviour will be high. When the cost of cooperation is high the perception of distinguishable tags will be low with a correspondingly low but non-zero level of cooperative behaviour. It would be interesting to see how ethnocentric behaviour, as exhibited by preferential donations to recognized kin in organisms which exhibit genetic kin discrimination, varies with the cost of donation. Organisms that exhibit kin discrimination occasionally donate to non-similar others; our results suggest that the degree to which this happens is sensitive to the cost of donation.  While overall donation rates will decrease when the cost of cooperation increases, when donations do occur they should be less specifically directed at individuals who share similar traits when the cost of donation is high.

\section*{Acknowledgements}
David McAvity would like to acknowledge financial support from a Sponsored Research grant from The Evergreen State College. We would like to thank anonymous reviewers for helpful suggestions.

\section*{Appendix}
\label{appendix}
\begin{figure}[!h]%
\begin{center}
\subfloat[]{\includegraphics[scale=1.0]{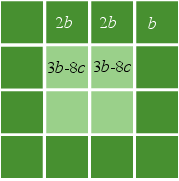}}\quad\quad\quad\quad%
\subfloat[]{\includegraphics[scale=1.0]{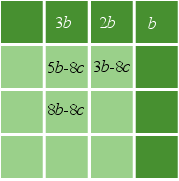}}
\end{center}
\caption{In (a) a 2x2 cluster of cooperators (light green) is
 surrounded by defectors (dark green).  In (b) a 3x3 cluster of cooperators (light green) is shown, and is assumed to be surrounded by defectors (dark green). Payoffs for agents important for establishing stability are shown.}%
 \label{fig:c-clusters}%
\end{figure}

In the appendix we give an explanation for transitions in cooperative behaviour that occur in imitation updating and how they depend on the number of tags. For imitation updating individuals adopt the strategy and tag of the neighbour with the highest payoff, so we just need to account for the payoffs that occur. Following  the work of Hautert {\em et.al.}~\citep{hauert2001fundamental}, we will examine fundamental clusters of cooperators and defectors to establish when such clusters are stable against invasion and when they can expand. First, a single cooperator surrounded by defectors will always be eliminated, since it will get no benefit but will incur costs. The smallest cluster of cooperators that can resist invasion by defectors is a $2\times2$ cluster. Such a cluster is illustrated in 
figure~\ref{fig:c-clusters}(a), with defectors indicated by dark green and cooperators by light green.  
Payoffs for agents important for establishing stability are shown.

This $2\times 2$ cluster is stable against invasion by defectors provided $3b-8c>2b \Rightarrow c/b<{1\over 8}$. In this case this is the same condition for the  $2\times 2$ cluster to expand. We next consider a $3\times3$ cluster of cooperators  in a sea of defectors as shown in figure~\ref{fig:c-clusters}(b). 
This  $3\times 3$ cluster is the smallest one which contains a cooperating agent completely surrounded by other cooperators. The presence of this agent increases the stability of this cluster since agents on the boundary of the cluster have a neighbouring cooperator who has a higher pay-off than they do. The cluster is stable against invasion provided $8b-8c>3b \Rightarrow c/b<{5\over 8}$. The condition for expanding out of the corner cell is $c/b<{1\over 8}$ as with $2\times 2$ clusters. However, the cluster can expand for higher costs, albeit more slowly, by invading defectors along the side provided $5b-8c>3b \Rightarrow c/b<{1\over 4}$. There is no advantage to having a larger cluster of cooperators in terms of stability or ability to expand. 

\begin{figure}[!h]%
\begin{center}
\subfloat[]{\includegraphics[scale=1.0]{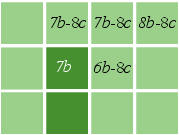}}\quad\quad\quad\quad
\subfloat[]{\includegraphics[scale=1.0]{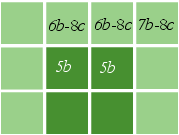}}
\end{center}
\caption{In (a) a pair of defectors (dark green) is surrounded by cooperators (light green). In (b) a $2\times 2$ cluster of defectors is surrounded by cooperators. Payoffs for agents that are important in determining the stability of clusters are indicated.}%
\label{fig:d-clusters}%
\end{figure}

We next consider configurations of defectors completely surrounded by cooperators. A single defector surround by cooperators will always have a higher pay-off and hence will expand into a $3\times 3$ cluster. The smallest cluster that would not always  expand regardless of cost is a neighbouring pair of defectors as shown in figure~\ref{fig:d-clusters}(a).

This cluster is always stable against invasion since each agent defecting is fitter than any of the cooperating neighbours. However, the condition for expansion is $7b>8b-8c  \Rightarrow  c/b > {1\over 8}$, in which case it would become a $3\times 4$ cluster. The smallest cluster of defectors that would be susceptible to invasion by cooperators is the $2\times 2$ cluster, for which each defector only receives payoff $5b$ as shown in~\ref{fig:d-clusters}(b). This cluster is invaded if $7b-8c>5b \Rightarrow c/b < {1\over 4}$. It expands if $5b>8b-8c \Rightarrow c/b > {3\over 8}$. Larger clusters do not gain a stability benefit, since additional defecting neighbours reduce rather then enhance fitness.  However, larger clusters are not more susceptible to invasion either. The fittest defecting agents are on the corner of a cluster, so agents most susceptible to invasion would occur on the straight edge border between cooperators and defectors. Such defecting agents would have at most 3 cooperating neighbours and would only get a pay-off of $3b$. The smallest cluster where a defecting agent and all its neighbours have this pay-off is a $5\times 2$ cluster. However, the neighbouring cooperators in this case get payoff $5b-8c$ so the condition for the invasion of this cluster is still $c/b < 1/4$.

 The results are summarized in the table below.
 
 \begin{tabular}{p{0.5in}|p{2.6in}|p{2.6in}}
  $c/b$ & Clusters of Cooperators  & Clusters of Defectors  \\ 
\hline 
[0,1/8) & $2\times 2$ clusters and larger expand. Smaller clusters are invaded. & Single defectors expand; pairs and strings of defectors with no more than two defecting neighbours persist. Larger clusters are invaded. \\ 
\hline 
(1/8,1/4)& $2\times2$ clusters and smaller are invaded. $3\times3$ clusters and larger expand.  & Pairs expand. Strings of defectors with two or defecting neighbours persist but do not expand. \\ 
\hline 
(1/4,3/8)& $3\times 3$ clusters persist but do not expand. & $2\times 2$ clusters persist but do not expand. Defectors in clusters with two or fewer defecting neighbours expand. \\ 
\hline 
(3/8,1/2) & $3\times 3$ clusters persist but do not expand.  & $2\times 2$ clusters expand, larger clusters expand only at corners with no more than 3 defecting neighbours. \\ 
\hline 
(1/2,5/8) & $3\times 3$ clusters persist but do not expand.  & Clusters of defectors expand at more complex corners with as many as 4 defecting neighbours. \\
\hline 
 (5/8,1] & All clusters are invaded. & Defectors fixate. \\ 
\end{tabular} 

Of course, due to the random nature of the initial conditions clusters may not initially have the configurations given in the table above. However, to the extend that the initial conditions result in a mixture of defectors and cooperators, defectors will tend to benefit from being surround by randomly dispersed cooperators. Cooperators would be at a disadvantage due to not being surround completely by other cooperators. We would thus expect defectors to initially expand into larger clusters. There would only be a few small islands of cooperators who happen to find themselves in a cluster due to the random initial conditions. At this stage, when considering the table above we would expect the cooperating clusters to expand if $c/b <1/4$, remain small, but stable for $1/4 \le c/b \le 5/8$ and be eliminated otherwise. Thus we expect a high rate of cooperation for $c/b \le 1/4$ but a low rate of cooperation otherwise.

When there is more than one tag, other strategies are possible. In particular, an {\em ethnocentric} strategy, whereby agents cooperate with others who share their own tag, but defect against others, is successful. Ethnocentric agents are successful because a cluster of ethnocentric agents surrounded by defecting agents with a different tag will always grow, regardless of cost, since they gain a fitness benefit from each other, but give no benefit to the defectors with a different tag. Two ethnocentric clusters with different tags are stable against invasion by the other. In addition, defectors of one tag type that are present along the boundary of two clusters of ethnocentric agents with different tags have a diminished ability to invade, so ethnocentric agents with one tag type help ethnocentric agents with different tag types resist invasion by defectors. We examine these claims quantitatively in what follows.  

\begin{figure}[!h]%
\begin{center}
\subfloat[]{\includegraphics[scale=1.0]{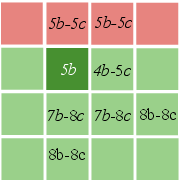}}\quad\quad
\subfloat[]{\includegraphics[scale=1.0]{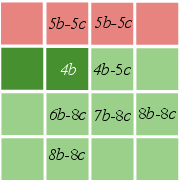 }}\quad\quad 
\subfloat[]{\includegraphics[scale=1.0]{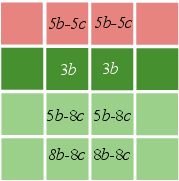}}
\end{center}
\caption{In (a) a single green defector is on the straight-line boundary between red and green clusters of ethnocentric agents. In (b) a pair of green defectors are on the straight-line boundary between red and green clusters of ethnocentric agents. In (c) the straight-line boundary between red and green clusters of ethnocentric agents is separated by a line of green defectors. Payoffs for agents that are important in determining the stability of clusters are indicated.}%
\label{fig:ethno-line-d}%
\end{figure}

Figure~\ref{fig:ethno-line-d}(a) shows a single green defector on the boundary between clusters of green and red ethnocentric agents. Without the presence of the red cluster the green defecting agent would invade the green ethnocentric agents, regardless of the cost. However, with this configuration it can only invade if  $5b>8b-8c \Rightarrow c/b>3/8$, in which case it would invade both types of clusters. Since a single green defector has a reduced benefit due to its proximity to red ethnocentric agents it can be invaded by green ethnocentric agents if $5b<7b-8c \Rightarrow c/b <1/4$. Larger clusters of defectors are even less successful at invading. For example, a pair of defectors, as shown in figure~\ref{fig:ethno-line-d}(b)  can only expand if $4b>8b-8\Rightarrow c/b>1/2$. The green pair of defectors will be eliminated by the green ethnocentric agents if $7b-8c>4b \Rightarrow  c/b<3/8$. A straight line of green defectors between the red and green ethnocentric clusters can only invade if $3b>8b-8c \Rightarrow c/b>5/8$. It will be invaded by the red ethnocentric agents if $5b-5c>3b\Rightarrow c/b<2/5$. We see that on some occasions the reduced fitness of the defectors allows ethnocentric agents with the same tag type to invade. On other occasions it is the ethnocentric agents with the other tag type that invade. Thus unlike others, with whom an ethnocentric agent defects, can indirectly enhance the stability of that agent's own cluster against defectors. It is in this way that diversity can enhance cooperation.

\begin{figure}[!h]%
\begin{center}
\subfloat[]{\includegraphics[scale=1.0]{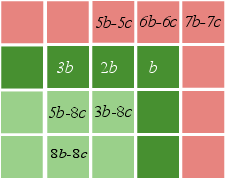}}\quad\quad%
\subfloat[]{\includegraphics[scale=1.0]{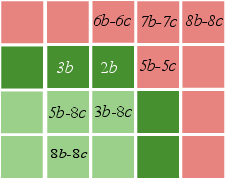}}
\quad\quad
\subfloat[]{\includegraphics[scale=1.0]{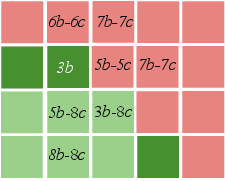}}
\end{center}
\caption{These figures show what happens at the corner of a green ethnocentric cluster surrounded by defectors in a sea of red ethnocentric agents. The initial situation is shown in (a). (b) Shows the configuration one generation later if $1/2<c/b<5/8$ and (c) shows the configuration one generation later if $2/5<c/b<1/2$. Payoffs for agents that are important in determining the stability of clusters are indicated.}%
\label{fig:ethno-corner-d}%
\end{figure}

Of course, not all boundaries between clusters are straight edge boundaries. Consider a cluster of green ethnocentric agents with a corner surround by a single layer of green defectors in a sea of red ethnocentric agents. A portion of this situation is illustrated in figure~\ref{fig:ethno-corner-d}(a). 

The red ethnocentric agents will invade the green defectors at the corner even for relatively high cost: $7b-7c>2b \Rightarrow c/b < 5/7$. The two agents neighbouring the corner one are also invaded if $6b-6c>3b \Rightarrow c/b<1/2$. The entire layer of defectors is invaded if $5b-5c>3b \Rightarrow c/b<2/5$, as with the straight line boundary in figure~\ref{fig:ethno-line-d}(c). Note that the defectors will invade both the green and red ethnocentric clusters along the sides unless $c/b < 5/8$. If we assume this condition then we arrive at the situation in figure~\ref{fig:ethno-corner-d}(b) for $1/2 <c/b < 5/8$. This configuration is then stable if $4/7 <c/b < 5/8$, but the defectors on the edge will be invaded if $1/2 <c/b < 4/7$, leading to the situation in figure~\ref{fig:ethno-corner-d}(c). We arrive at this situation directly from the initial configuration if $2/5<c/b<1/2$. The green defector on the leading edge will then be invaded if $c/b < 4/7$, eliminating the entire line of defectors in a step by step process.  

Many other configurations  of clusters are possible, each with its own stability and invasion criteria. However, the interesting  transition in behaviour occurs in the interval $2/5<c/b<1/2$. Below this interval large clusters of defectors are eliminated by ethnocentric agents with different tag types. Above this interval defecting agents can resist invasion and are able to expand in most situations.  The important point is that defecting agents that invade a cluster of like tagged agents will eventually expand until they reach a boundary with ethnocentric agents that have a different tag. When they do, they may be susceptible to invasion by that ethnocentric cluster, depending on the cost of cooperation. The consequence is that the frequency of ethnocentric agents in the population grows, and the overall cooperation rate is higher than it would be if there were no conditional tag based strategies. We expect to see a transition from highly cooperative behaviour to low cooperative behaviour in the interval $2/5<c/b<1/2$. This is higher than the similar transition that occurs at $c/b=1/4$ when there are no tag-based conditional strategies.

\section*{References}
\label{references}

\bibliographystyle{unsrt}
\bibliography{evolution}

\begin{thebibliography}{10}

\bibitem{darwin1859origin}
C.~Darwin.
\newblock {\em On the origin of species}.
\newblock Harvard University Press, 1859.

\bibitem{smith1997major}
J.M. Smith and E.~Szathm{\'a}ry.
\newblock {\em The major transitions in evolution}.
\newblock Oxford University Press, USA, 1997.

\bibitem{west2006social}
S.A. West, A.S. Griffin, A.~Gardner, and S.P. Diggle.
\newblock Social evolution theory for microorganisms.
\newblock {\em Nature Reviews Microbiology}, 4(8):597--607, 2006.

\bibitem{crespi2001evolution}
B.J. Crespi.
\newblock The evolution of social behavior in microorganisms.
\newblock {\em TRENDS in Ecology \& Evolution}, 16(4), 2001.

\bibitem{clutton2000individual}
TH~Clutton-Brock, PNM Brotherton, MJ~O'Riain, AS~Griffin, D.~Gaynor, L.~Sharpe,
  R.~Kansky, MB~Manser, and GM~McIlrath.
\newblock Individual contributions to babysitting in a cooperative mongoose,
  suricata suricatta.
\newblock {\em Proceedings of the Royal Society of London. Series B: Biological
  Sciences}, 267(1440):301--305, 2000.

\bibitem{sharp2005learned}
S.P. Sharp, A.~McGowan, M.J. Wood, and B.J. Hatchwell.
\newblock Learned kin recognition cues in a social bird.
\newblock {\em Nature}, 434(7037):1127--1130, 2005.

\bibitem{maynard-smith1974evolutionarygames}
J.~Maynard~Smith.
\newblock The theory of games and the evolution of animal conflicts.
\newblock {\em Journal of theoretical biology}, 47:209--221, 1974.

\bibitem{nowak1992evolutionary}
M.A. Nowak and R.M. May.
\newblock {Evolutionary games and spatial chaos}.
\newblock {\em Nature}, 359(6398):826--829, 1992.

\bibitem{nowak1994more}
M.A. Nowak, S.~Bonhoeffer, and R.M. May.
\newblock {More spatial games}.
\newblock {\em International Journal of Bifurcation and Chaos}, 4:33--33, 1994.

\bibitem{szabo1998evolutionary}
G.~Szab{\'o} and C.~T{\H{o}}ke.
\newblock {Evolutionary prisoner's dilemma game on a square lattice}.
\newblock {\em Physical Review E}, 58(1):69, 1998.

\bibitem{schweitzer2002evolution}
F.~Schweitzer, L.~Behera, and H.~M\"{u}hlenbein.
\newblock {Evolution of cooperation in a spatial prisoner's dilemma}.
\newblock {\em Advances in Complex Systems}, 5(02):269--299, 2002.

\bibitem{langer2008spatial}
P.~Langer, M.A. Nowak, and C.~Hauert.
\newblock {Spatial invasion of cooperation}.
\newblock {\em Journal of theoretical biology}, 250(4):634--641, 2008.

\bibitem{wright1943isolation}
S.~Wright.
\newblock Isolation by distance.
\newblock {\em Genetics}, 28(2):114, 1943.

\bibitem{taylor1992altruism}
PD~Taylor.
\newblock Altruism in viscous populations—an inclusive fitness model.
\newblock {\em Evolutionary Ecology}, 6(4):352--356, 1992.

\bibitem{lehmann2006population}
L.~Lehmann, N.~Perrin, and F.~Rousset.
\newblock Population demography and the evolution of helping behaviors.
\newblock {\em Evolution}, 60(6):1137--1151, 2006.

\bibitem{rousset2000theoretical}
F.~Rousset and S.~Billiard.
\newblock A theoretical basis for measures of kin selection in subdivided
  populations: finite populations and localized dispersal.
\newblock {\em Journal of Evolutionary Biology}, 13(5):814--825, 2000.

\bibitem{huberman1993evolutionary}
B.A. Huberman and N.S. Glance.
\newblock Evolutionary games and computer simulations.
\newblock {\em Proceedings of the national academy of sciences of the United
  States of America}, 90(16):7716, 1993.

\bibitem{ohtsuki2006replicator}
H.~Ohtsuki and M.A. Nowak.
\newblock The replicator equation on graphs.
\newblock {\em Journal of theoretical biology}, 243(1):86--97, 2006.

\bibitem{ohtsuki2008evolutionary}
H.~Ohtsuki and M.A. Nowak.
\newblock Evolutionary stability on graphs.
\newblock {\em Journal of theoretical biology}, 251(4):698--707, 2008.

\bibitem{hamilton1964genetical}
W.D. Hamilton.
\newblock {The genetical evolution of social behaviour. I}.
\newblock {\em Journal of theoretical biology}, 7(1):1--16, 1964.

\bibitem{dawkins1976selfish}
R.~Dawkins.
\newblock {\em The selfish gene}.
\newblock Oxford University, 1976.

\bibitem{queller2003single}
D.C. Queller, E.~Ponte, S.~Bozzaro, and J.E. Strassmann.
\newblock Single-gene greenbeard effects in the social amoeba dictyostelium
  discoideum.
\newblock {\em Science}, 299(5603):105, 2003.

\bibitem{dawkins1982extended}
R.~Dawkins.
\newblock {\em The Extended Phenotype}.
\newblock Oxford University Press, 1982.

\bibitem{crozier}
R.H. Crozier.
\newblock Genetic clonal recognition abilities in marine invertebrates must be
  maintained by selection for something else.
\newblock {\em Evolution}, 40:1100--1101, 1986.

\bibitem{rousset2007constraints}
F.~Rousset and D.~Roze.
\newblock Constraints on the origin and maintenance of genetic kin recognition.
\newblock {\em Evolution}, 61(10):2320--2330, 2007.

\bibitem{jansen2006altruism}
V.A.A. Jansen and M.~Van~Baalen.
\newblock {Altruism through beard chromodynamics}.
\newblock {\em Nature}, 440(7084):663--666, 2006.

\bibitem{gardner2007social}
A.~Gardner and S.A. West.
\newblock Social evolution: the decline and fall of genetic kin recognition.
\newblock {\em Current Biology}, 17(18):R810--R812, 2007.

\bibitem{riolo2001evolution}
R.L. Riolo, M.D. Cohen, and R.~Axelrod.
\newblock {Evolution of cooperation without reciprocity}.
\newblock {\em Nature}, 414(6862):441--443, 2001.

\bibitem{sigmund2001tides}
K.~Sigmund and M.A. Nowak.
\newblock {Tides of tolerance.}
\newblock {\em Nature}, 414(6862):403--405, 2001.

\bibitem{traulsen2003minimal}
A.~Traulsen and H.G. Schuster.
\newblock {Minimal model for tag-based cooperation}.
\newblock {\em Physical Review E}, 68(4):046129, 2003.

\bibitem{axelrod2004altruism}
R.~Axelrod, R.A. Hammond, and A.~Grafen.
\newblock {Altruism via kin-selection strategies that rely on arbitrary tags
  with which they coevolve}.
\newblock {\em Evolution}, 58(8):1833--1838, 2004.

\bibitem{hammond2006evolution}
R.A. Hammond and R.~Axelrod.
\newblock {Evolution of contingent altruism when cooperation is expensive}.
\newblock {\em Theoretical population biology}, 69(3):333--338, 2006.

\bibitem{roberts2002Similarity}
G.~Roberts and T.N. Sherratt.
\newblock Does similarity breed cooperation?
\newblock {\em Nature}, 418(6897):499--500, 2002.

\bibitem{traulsen2007chromodynamics}
A.~Traulsen and M.A. Nowak.
\newblock Chromodynamics of cooperation in finite populations.
\newblock {\em PLoS One}, 2(3):e270, 2007.

\bibitem{hammond2006ethnocentrism}
R.A. Hammond and R.~Axelrod.
\newblock The evolution of ethnocentrism.
\newblock {\em Journal of Conflict Resolution}, 50(6):926, 2006.

\bibitem{mateo2000kin}
J.M. Mateo and R.E. Johnston.
\newblock Kin recognition and the ‘armpit effect’: evidence of
  self--referent phenotype matching.
\newblock {\em Proceedings of the Royal Society of London. Series B: Biological
  Sciences}, 267(1444):695--700, 2000.

\bibitem{szabo2002phase}
G.~Szab{\'o} and C.~Hauert.
\newblock Phase transitions and volunteering in spatial public goods games.
\newblock {\em Physical review letters}, 89(11):118101, 2002.

\bibitem{traulsen2006stochastic}
A.~Traulsen, M.A. Nowak, and J.M. Pacheco.
\newblock Stochastic dynamics of invasion and fixation.
\newblock {\em Physical Review E}, 74(1):011909, 2006.

\bibitem{hauert2001fundamental}
C.~Hauert.
\newblock {Fundamental clusters in spatial 2$\times$ 2 games}.
\newblock {\em Proceedings of the Royal Society of London. Series B: Biological
  Sciences}, 268(1468):761, 2001.

\bibitem{killingback1996spatial}
T.~Killingback and M.~Doebeli.
\newblock Spatial evolutionary game theory: Hawks and doves revisited.
\newblock {\em Proceedings of the Royal Society of London. Series B: Biological
  Sciences}, 263(1374):1135, 1996.

\bibitem{shultz2008stages}
T.R. Shultz, M.~Hartshorn, and R.A. Hammond.
\newblock Stages in the evolution of ethnocentrism.
\newblock In {\em Proceedings of the 30th annual conference of the cognitive
  science society}, pages 1244--1249, 2008.

\bibitem{ohtsuki2006simple}
H.~Ohtsuki, C.~Hauert, E.~Lieberman, and M.A. Nowak.
\newblock A simple rule for the evolution of cooperation on graphs and social
  networks.
\newblock {\em Nature}, 441(7092):502--505, 2006.

\end{thebibliography}
\end{document}